\documentclass[aps,prl,twocolumn,amssymb,amsfonts,nobibnotes,floatfix,nofootinbib,superscriptaddress,longbibliography]{revtex4-1}
\usepackage[utf8]{inputenc}
\pdfoutput=1
\usepackage{color}
\usepackage{multirow}
\usepackage{dcolumn}
\usepackage{float}
\usepackage{xcolor}
\usepackage{amsmath,amsthm,amsfonts,amssymb}
\usepackage{times}
\usepackage[normalem]{ulem}
\usepackage{commath}
\usepackage{physics}
\usepackage{bm}
\usepackage[pdftex]{graphicx}
\usepackage[colorlinks=false]{hyperref}

\newcommand{\panelref}[2]{({\normalsize{#1}\:\!{\footnotesize #2})}}

\begin{document}

\title{Ubiquitous quantum scarring does not prevent ergodicity\\{\normalfont \emph{ Published version: \href{\doibase 10.1038/s41467-021-21123-5}{10.1038/s41467-021-21123-5}}}}

\author{Sa\'ul Pilatowsky-Cameo}
\author{David Villase\~nor}
\affiliation{Instituto de Ciencias Nucleares, Universidad Nacional Aut\'onoma de M\'exico, Apdo. Postal 70-543, C.P. 04510, CDMX, Mexico}
\author{Miguel~A.~Bastarrachea-Magnani}
\affiliation{Department of Physics and Astronomy, Aarhus University, Ny Munkegade, DK-8000 Aarhus C, Denmark}
\affiliation{Departamento de F\'isica, Universidad Aut\'onoma Metropolitana-Iztapalapa, San Rafael Atlixco 186, C.P. 09340, CDMX, Mexico}
\author{Sergio Lerma-Hern\'andez}
\affiliation{Facultad de F\'isica, Universidad Veracruzana, Circuito Aguirre Beltr\'an s/n, Xalapa, Veracruz 91000, Mexico}
\author{Lea~F.~Santos} 
\email[email: ]{lsantos2@yu.edu, hirsch@nucleares.unam.mx} %leave both emails here, so there is only one footnote with both emails as required

\affiliation{Department of Physics, Yeshiva University, New York, New York 10016, USA}
\author{Jorge G. Hirsch}
\email[email: ]{lsantos2@yu.edu, hirsch@nucleares.unam.mx} %leave both emails here, so there is only one footnote with both emails as required

\affiliation{Instituto de Ciencias Nucleares, Universidad Nacional Aut\'onoma de M\'exico, Apdo. Postal 70-543, C.P. 04510, CDMX, Mexico}

%%%%%%%%%%%%%% ABSTRACT %%%%%%%%%%%%%%%%f
   
\begin{abstract}
{\bf In a classically chaotic system that is ergodic, any trajectory will be arbitrarily close to any point of the available phase space after a long time, filling it uniformly. Using Born's rules to connect quantum states with probabilities, one might then expect that all quantum states in the chaotic regime should be uniformly distributed in phase space. This simplified picture was shaken by the discovery of quantum scarring, where some eigenstates are concentrated along unstable periodic orbits.  Despite of that, it is widely accepted that most eigenstates of chaotic models are indeed ergodic. Our results show instead that all eigenstates of the chaotic Dicke model are actually scarred. They also show that even the most random states of this interacting atom-photon system never occupy more than half of the available phase space. Quantum ergodicity is achievable only as an ensemble property, after temporal averages are performed.}

\end{abstract}

%%%%%%%%%%%%%%%%%%%%%%%%%%%%%%%%%%%%%%%%
\maketitle
%%%%%%%%%%%%%%%%%%%%%%%%%%%%%%%%%%%%%%%%

%%%%%%%%%%%%%%%% INTRODUCTION %%%%%%%%%%%%%%%%
A striking feature of the quantum-classical correspondence not recognized in the early days of the quantum theory is the repercussion that measure-zero structures of the classical phase space may have in the quantum domain. A recent example is the effect of unstable fixed points, that cause the exponentially fast scrambling of quantum information in both integrable and chaotic quantum systems~\cite{Pappalardi2018,Pilatowsky2019,Hummel2019,Tianrui2020,Hashimoto2020}. Another better known example is the phenomenon of quantum scarring~\cite{Heller1984,Heller1991,Kaplan1999}. As a parameter of a classical system is varied and it transits from a regular to a chaotic regime, periodic orbits that may be present in the phase space change from stable to unstable. These classical unstable periodic orbits  can get imprinted in the quantum states as regions of concentrated large amplitudes known as quantum scars. Even though the phase space may be densely filled with unstable periodic orbits, they are still of measure zero, which explains why it took until the works by Gutzwiller~\cite{Gutzwiller1971} for their importance in the quantum chaotic dynamics to be finally recognized. 

Quantum scarring was initially observed in the Bunimovich stadium billiard~\cite{McDonald1979} and soon in various other one-body systems~\cite{StockmannBook,Kus1991,Ariano1992} giving rise to a new line of research in the field of quantum chaos~\cite{Heller1987,Bogomolny1988,Berry1989,Agam1993,Bohigas1993,Wintgen1989,Muller1994,%Kaplan1998,
Kaplan1999,Wisniacki2006,Revuelta2020}. The recent experimental observation of long-lived oscillations in chains of Rydberg atoms~\cite{Bernien2017}, associated with what is now called ``many-body quantum scars'', has caused a new wave of fascination with the phenomenon of quantum scarring~\cite{Turner2018,Ho2019,Choi2019,Shriya2019,Turner2020Arxiv}. While the interest in many-body quantum scars lies in their potential as resources to manipulate and store quantum information, a direct relationship between them and possible structures in the classical phase space has not yet been established. 

Halfway between one-body and many-body models, one finds systems such as two-dimensional harmonic oscillators and the Dicke model~\cite{Dicke1954}, 
where quantum scars have also been observed \cite{Deaguiar1992,Bakemeier2013}. In the first case, the model is not fully chaotic and scarring can be understood as an extension of the regular orbits \cite{Keski2019,KeskiHeller2019}. The Dicke model, on the other hand, has a region of strong chaos, where the Lyapunov exponents are positive and the level statistics agrees with the predictions from random matrix theory~\cite{Chavez2016}. The model describes a large number of two-level atoms that interact collectively with a quantized radiation field and was first introduced to explain the phenomenon of superradiance~\cite{Hepp1973a,%Wang1973,Garraway2011,
Kirton2019}. It has been studied experimentally with cavity assisted Raman transitions~\cite{Baden2014,Zhang2018}, trapped ions~\cite{Cohn2018,Safavi2018}, and superconducting circuits~\cite{Jaako2016}.

In this work, we investigate the intricate relationship between quantum scarring and phase-space localization in the superradiant phase of the Dicke model. Even though both phenomena are often treated on an equal footing, the connection is rather subtle. Scarring refers to structures that resemble periodic orbits in the phase-space distribution of quantum eigenstates, while phase-space localization implies that a state exhibits a low degree of spreading in the phase space. Here, we demonstrate that scarring does not necessarily imply significant phase-space localization.

In systems studied before, scarred eigenstates were thought to be a fraction of the total number of eigenstates. In contrast to that, we show that deep in the chaotic regime of the Dicke model, all eigenstates are scarred. Their phase-space probability distributions always display structures that can be traced back to periodic orbits in the classical limit. Yet, we find eigenstates that are highly localized in phase space and eigenstates that are nearly as much spread out as random states, although none of them, including the random states, can cover more than approximately half of the available phase space. 

In addition to the analysis of quantum scarring and phase-space localization, we also provide a definition of quantum ergodicity. This is done using a measure that we introduce to quantify the level of localization of quantum states in the phase space. We say that a quantum state is ergodic if its infinite-time average leads to full delocalization.  Under this definition, stationary quantum states are never ergodic, while random states are, and coherent states lie somewhere in between.

%%%%%%%%%%%%%%%%%%%%%%%%%%%%%%%%%%%%%%%%%%%%%%%%%%%%%
%%%%%%%%%%%%%%%%%%%% FIGURE 1 %%%%%%%%%%%%%%%%%%%%%%%%%%%
\begin{figure*}[t]
\centering
\begin{tabular}{l}
\includegraphics[width=1\textwidth]{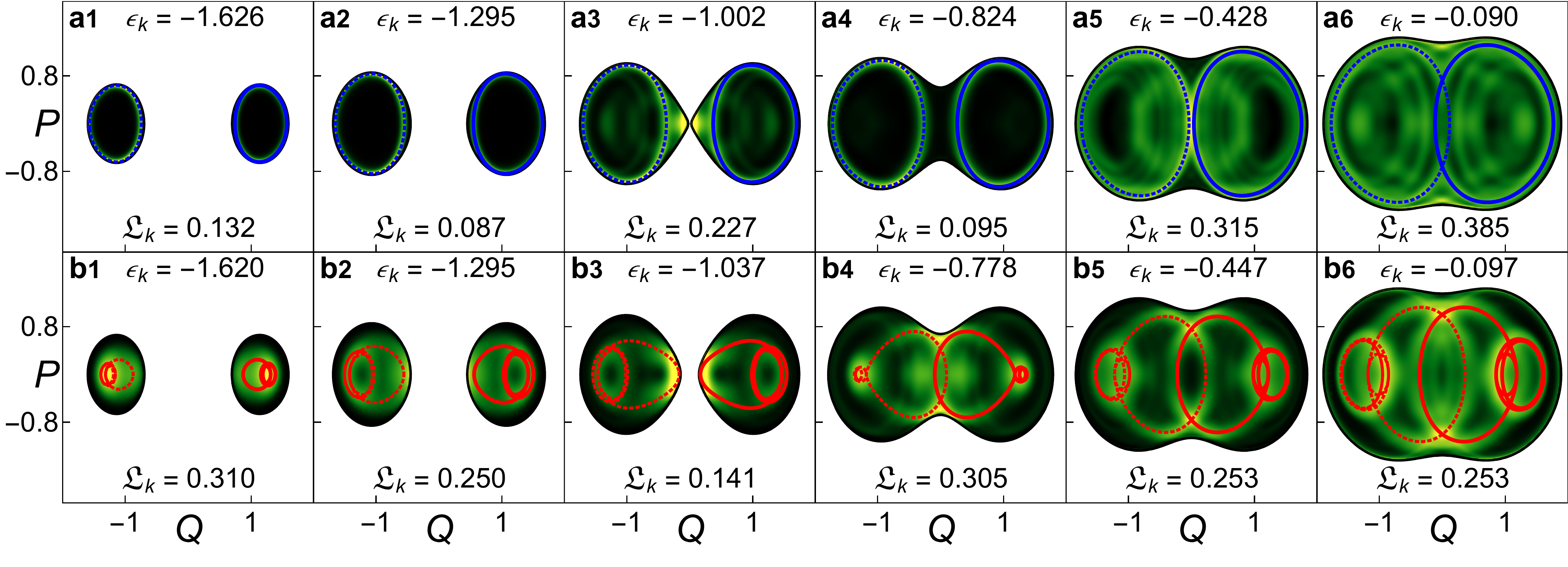} 
\end{tabular}
\caption{\textbf{Classical periodic orbits and scars in the Husimi projection of eigenstates.} \textbf{\panelref{a}{1}-\panelref{a}{6}} [\textbf{\panelref{b}{1}-\panelref{b}{6}}] Projected Husimi distribution $\widetilde{\mathcal{Q}}_k(Q,P)$ superposed by periodic orbits from the family ${\cal A}$ (blue lines) [family ${\cal B}$ (red lines)]. Dashed lines mark the mirror image (in $Q$ and $q$) of the periodic orbits drawn with solid lines. The mirror images are also periodic orbits due to the parity symmetry of the  Hamiltonian. In the projected Husimi distributions, lighter colors indicate higher concentrations, while black corresponds to zero. The values of the energy $\epsilon_k$ and localization measure $\mathfrak{L}_k$ of each eigenstate $k$ are indicated in the panels. Energies larger than $-0.8$ are in the chaotic region. The system size is $j=30$.  
}
\label{fig01}
\end{figure*}
%%%%%%%%%%%%%%%%%%%%%%%%%%%%%%%%%%%%%%%%%%%%%%%%%%%%%
%%%%%%%%%%%%%%%%%%%%%%%%%%%%%%%%%%%%%%%%%%%%%%%%%%%%%

%%%%%%%%%%%%%%% MODEL and CHAOS %%%%%%%%%%%%%%%%%%%%%%%%%
%\vskip 1 cm
\vskip 0.1 cm
{\bf\large Results}\\
{\bf Dicke model and chaos.} The Hamiltonian of the Dicke model is written as
\begin{equation}
\label{eqn:qua_hamiltonian}
\hat{H}_{D}=\omega\hat{a}^{\dagger}\hat{a}+\omega_{0}\hat{J}_{z}+\frac{2\gamma}{\sqrt{\mathcal{N}}}\hat{J}_{x}(\hat{a}^{\dagger}+\hat{a}),
\end{equation}
where $\hbar=1$. It describes $\mathcal{N}$ two-level atoms with atomic transition frequency $\omega_{0}$ interacting with a single mode of the electromagnetic field with radiation frequency $\omega$. In the equation above, $\hat{a}$ ($\hat{a}^{\dagger}$) is the bosonic annihilation (creation) operator of the field mode,  $\hat{J}_{x,y,z}=\frac{1}{2}\sum_{k=1}^{\mathcal{N}}\hat{\sigma}_{x,y,z}^{k}$ are the collective pseudo-spin operators, with  $\hat{\sigma}_{x,y,z}$ being the Pauli matrices, and $\gamma$ is the atom-field coupling strength. 

The eigenvalues $j(j+1)$ of the squared total spin operator $\hat{\bold{J}}^{2}=\hat{J}_{x}^{2}+\hat{J}_{y}^{2}+\hat{J}_{z}^{2}$ specify the different invariant subspaces of the model. We use the symmetric atomic subspace defined by the maximum pseudo-spin $j=\mathcal{N}/2$, which includes the ground state. When the Dicke model reaches the critical value $\gamma_{c}=\sqrt{\omega\omega_{0}}/2$, it goes from a normal phase ($\gamma<\gamma_c$) to a superradiant phase ($\gamma>\gamma_c$). Our studies are done in the superradiant phase, $\gamma=2\gamma_c$, and we choose $\omega=\omega_0=1$. The rescaled energies are denoted by $\epsilon=E/j$. For the selected parameters, $\epsilon_\text{GS}=-2.125$ is the ground-state energy. 

The classical Hamiltonian, $h_\text{cl}(\bold{x})$ in the coordinates $\bold{x}=(q,p;Q,P)$, is obtained by calculating the expectation value of the quantum Hamiltonian under the  product of bosonic Glauber  and  pseudo-spin Bloch coherent states  $|\bold{x}\rangle=|q,p\rangle\otimes~|Q,P\rangle$ (see Methods) and dividing it by $j$. The effective Planck constant $\hbar_{\text{eff}}=1/j$~\cite{Ribeiro2006} determines the resolution of the quantum states on the four dimensional phase space $\mathcal{M}$. We are able to work with large system sizes ($j\sim 100$ and Hilbert space dimensions $D\sim 6\times 10^4$), due to the use of an efficient basis that guarantees the convergence of a broad range of eigenvalues and eigenstates~\cite{Chen2008,Bastarrachea2014PSa%,Bastarrachea2014PSb
} (see Methods). 
The Dicke model displays regular and chaotic behavior. For the Hamiltonian parameters selected in this work, the system is in the strong-coupling hard-chaos regime for $\epsilon > -0.8$ (see Supplementary Note 1).  

%%%%%%%%%%%%%% 	EIGENSTATES %%%%%%%%%%%%%%%%
\vskip 0.2 cm
{\bf Quantum scarring.} 
The (unnormalized) Husimi function of a state $\hat{\rho}$ is defined as $\mathcal{Q}_{\hat{\rho}}(\bold{x})= \langle \bold{x} | \hat{\rho} |\bold{x} \rangle$, which is the expectation value of the density matrix over the coherent state $\ket{\bold{x}}$ centered at $\bold{x}$. This function is used to visualize how the state $\hat{\rho}$ is distributed in the phase space. 
Quantum scars are localized around the classical periodic orbits in an energy shell of the phase space. To visualize the scars, we consider the Husimi projection over the classical energy shell at $\epsilon$, 
\begin{equation}
\label{eq:projQdef}
\widetilde{\mathcal{Q}}_{\epsilon,\hat{\rho}}(Q,P)= \iint \dif q \dif p \, \delta \big(\epsilon - h_\text{cl}(\bold{x}) \big) \mathcal{Q}_{\hat{\rho}}(\bold{x}).
\end{equation}
By integrating out the bosonic variables $(q,p)$, the remaining function can be compared with the projection of the classical periodic orbits on the  plane  of atomic variables $(Q,P)$.

Identifying all periodic orbits that generate the scars of a quantum system is extremely challenging. We were able to identify two families of periodic orbits for the Dicke model, which we denote by family ${\cal A}$ and family ${\cal B}$ \cite{Pilatowsky2021}. By calculating the overlap of the eigenstates with tubular phase-space distributions located around these orbits \cite{Kaplan1999}, we selected twelve eigenstates $\hat{\rho}_k=\dyad{E_k}$ scarred by those two different families. In Fig.~\ref{fig01} we plot their Husimi projections $\widetilde{\mathcal{Q}}_k=\widetilde{\mathcal{Q}}_{\epsilon_k,\hat{\rho}_k}$ at $\epsilon_{k}=E_k/j$ along with the corresponding periodic orbit of each family. Family ${\cal A}$ [solid blue line in Fig.~\ref{fig01}~\panelref{a}{1}-\panelref{a}{6}] contains the periodic orbits of lowest period of the Dicke model, which emanate from one of the two normal modes  around  a stable stationary point at the ground-state energy. Family ${\cal B}$ [solid red line in Fig.~\ref{fig01}~\panelref{b}{1}-\panelref{b}{6}] arises from the other normal mode  around the same point. Scarring is clearly visible in all panels of Fig. \ref{fig01}. The quantum states are highly concentrated around the classical periodic orbits. This happens even in the chaotic region of high excitation energy, where the classical dynamics is ergodic, as seen in Figs.~\ref{fig01}~\panelref{a}{5}, \panelref{a}{6}, \panelref{b}{5}, and \panelref{b}{6}. Notice that the eigenstates may be scarred by more than one periodic orbit. In fact, as we showed quantitatively in ref. ~\cite{Pilatowsky2021} after introducing a measure of scarring, an eigenstate may even be scarred by periodic orbits of different families. This means that different eigenstates exhibit different degrees of scarring.

It is evident from Fig.~\ref{fig01} that the degree of delocalization of the eigenstates in phase space also varies. The Husimi distribution of the eigenstates in Fig.~\ref{fig01}~\panelref{a}{5} and \panelref{a}{6}, for instance, is not entirely confined to the two periodic orbits drawn in blue. This contrasts with the high density concentration that the eigenstate in Fig.~\ref{fig01}~\panelref{a}{4} shows around the plotted unstable periodic orbits. To quantify these differences, we introduce a measure of the degree of localization of a quantum state in the classical energy shell. 

\vskip 0.2 cm 
{\bf Scarring vs. phase-space localization.}
To measure the localization of a state in a Hilbert-space basis indexed by some letter $n$, the most commonly used quantity is the participation ratio $P_R$ \cite{Edwards1972,Gnutzmann2001,Romera2012}. Its inverse is given by $P_R^{-1}=\sum_n \mathcal{P}_n^2$, where $\mathcal{P}_n$ is the probability of finding the state in the $n$'th basis vector. By analogy, we introduce a measure of localization in phase space that employs as basis the overcomplete set of coherent states within a single energy shell $\mathcal{M}_\epsilon=\{\bold{x}=(q,p;Q,P)\mid h_\text{cl}(\bold{x})=\epsilon \}$, so that we replace the sum $\sum_k$ with a three-dimensional surface integral $\int_{\mathcal{M_\epsilon}}\!\!\dif\bold{s}$ over $\mathcal{M}_\epsilon$. For a given $\bold{x} \in \mathcal{M}_\epsilon$, the probability $\mathcal{P}_{\bold{x}}$ of finding the state $\hat{\rho}$ in the coherent state $\ket{\bold{x}}$ is given by the Husimi function $\mathcal{Q}_{\hat{\rho}}(\bold{x})$. With these replacements, we finally obtain a measure of phase-space localization $\mathfrak{L}(\epsilon,\hat{\rho})$ given by
\begin{equation}
\label{eqn:LocMeasDef}
\mathfrak{L}(\epsilon,\hat{\rho})^{-1} = \frac{1}{N} \int_\mathcal{M_\epsilon}  \dif \bold{s} \, \mathcal{Q}_{\hat{\rho}}^{2}(\bold{x}),
\end{equation}
where $N=\big(\int_{\mathcal{M}_\epsilon} \dif \bold{s} \, \mathcal{Q}_{\hat{\rho}}(\bold{x})\big)^2/\mathcal{V}(\epsilon)$ is a normalization constant and $\mathcal{V}(\epsilon) =  \int_{\mathcal{M}_\epsilon}  \dif \bold{s}$ is the volume of $\mathcal{M}_\epsilon$ (see Methods).

The measure $\mathfrak{L}(\epsilon,\hat{\rho})$ is an energy-restricted second moment of the Husimi function~\cite{Sugita2002}. It is related to the second-order R\'enyi-Wehrl entropy~\cite{Gnutzmann2001}, which, in turn, was shown for the Dicke model~\cite{Wang2020} to be linearly related to the first-order R\'enyi-Wehrl entropy~\cite{Wehrl1979}.

%%%%%%%%%%%%%%%%%%%%%%%%%%%%%%%%%%%%%%%%%%%%%%%%%%%%%
%%%%%%%%%%%%%%%%%%%% FIGURE 2 %%%%%%%%%%%%%%%%%%%%%%%%%%%
\begin{figure*}[ht]
\includegraphics[width=1\textwidth]{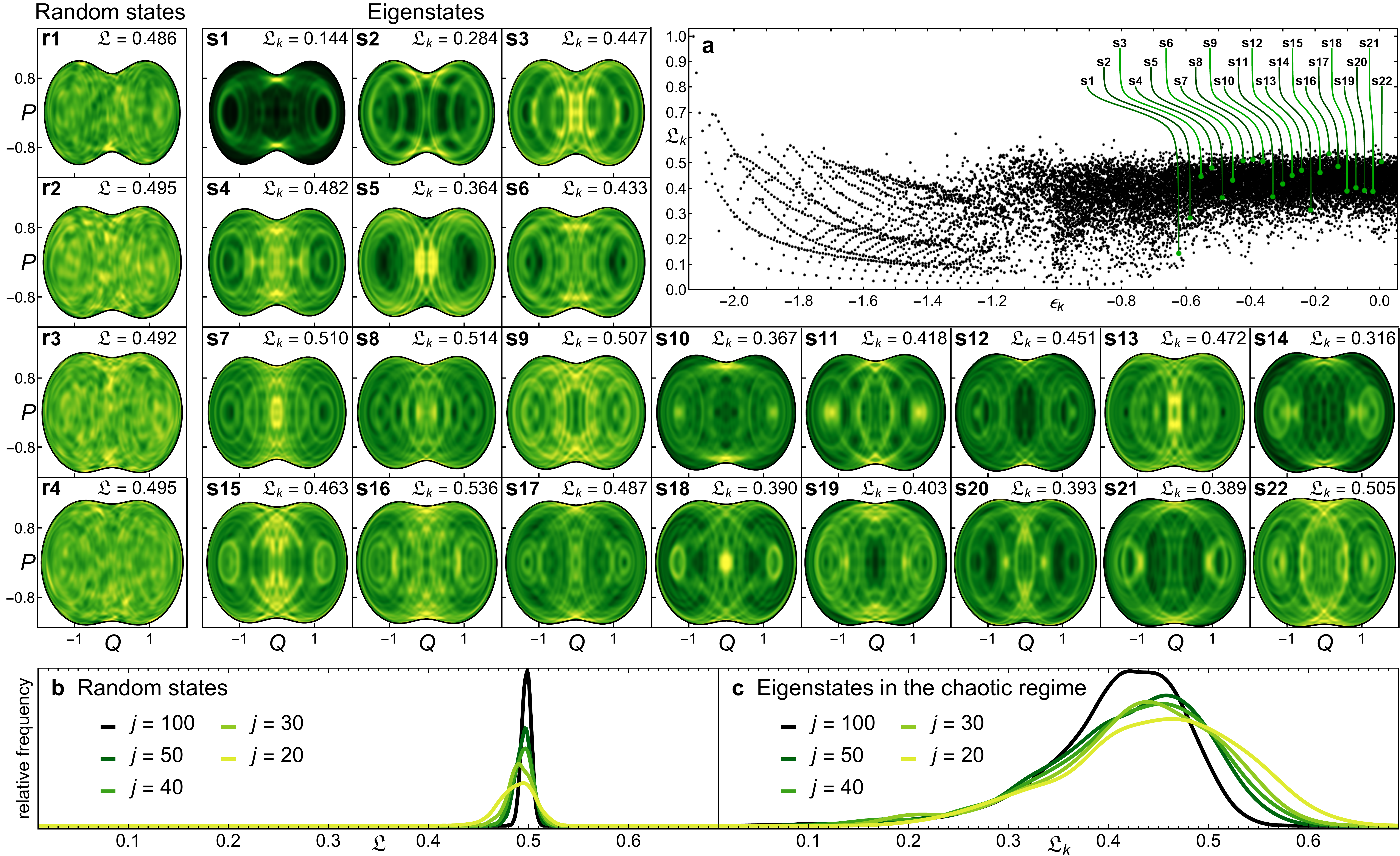}
\caption{\textbf{Husimi projection and localization measure of eigenstates and random states.} \textbf{(a)} Localization measure $\mathfrak{L}_{k}$ as a function of energy for the first 17,000 eigenstates with $\epsilon_k \in [\epsilon_{GS},0.1]$, $j=100$.   \textbf{\panelref{s}{1}-\panelref{s}{22}} Projected Husimi distributions for 22 eigenstates selected every $400$ values of $k$, starting at $k=7700$ in \panelref{s}{1} up to $k=16500$ in \panelref{s}{22}, $j=100$. Lighter colors indicate higher concentrations, while black corresponds to zero. \textbf{\panelref{r}{1}-\panelref{r}{4}} Projected Husimi distributions for random states of energy width $\sigma=0.3$ centered at $\epsilon=-0.6$ in \panelref{r}{1},  $\epsilon=-0.4$ in \panelref{r}{2}, $\epsilon=-0.2$ in \panelref{r}{3}, and $\epsilon=-0.1$ in \panelref{r}{4}, $j=100$. Lighter colors indicate higher concentrations, while black corresponds to zero. \textbf{(b)}  Distribution of $\mathfrak{L}$ for 20,000 random states of width $\sigma=0.3$ centered at $\epsilon=-0.5$. \textbf{(c)} Distribution of $\mathfrak{L}_k$ for the eigenstates in the chaotic region with $\epsilon_k\in [-0.8,0]$. The system sizes for (b) and (c) are indicated in the panels.
}
\label{fig02}
\end{figure*}
%%%%%%%%%%%%%%%%%%%%%%%%%%%%%%%%%%%%%%%%%%%%%%%%%%%%%
%%%%%%%%%%%%%%%%%%%%%%%%%%%%%%%%%%%%%%%%%%%%%%%%%%%%%

The value of $\mathfrak{L}(\epsilon,\hat{\rho})$ indicates the fraction of the classical energy shell at $\epsilon$ that is covered by the state $\hat{\rho}$. It varies from its minimum value $\mathfrak{L}(\epsilon,\hat{\rho})\sim (2\pi \hbar_\text{eff})^2/\mathcal{V}(\epsilon)$, which indicates maximum localization, to $\mathfrak{L}(\epsilon,\hat{\rho})=1$, which implies complete delocalization over the energy shell. The former 
occurs for coherent states, and the latter happens if $\mathcal{Q}_{\hat{\rho}}(\bold{x})$ is a constant for all $\bold{x} \in \mathcal{M}_{\epsilon}$, in which case the projection $\widetilde{\mathcal{Q}}_{\epsilon,\hat{\rho}}(Q,P)$ is also constant for the allowed values of $Q$ and $P$.

All eigenstates in Fig.~\ref{fig01} have values of  $\mathfrak{L}_{k} =\mathfrak{L}(\epsilon_k,\hat{\rho}_k)$ 
below $1/2$. For the eigenstates  in Fig.~\ref{fig01}~\panelref{a}{1}, \panelref{a}{2}, \panelref{a}{4}, \panelref{b}{3}, these values are very small, since the eigenstates are almost entirely localized around the plotted periodic orbits. The value of $\mathfrak{L}_{k}$ is larger in Fig.~\ref{fig01}~\panelref{a}{3}, because at the center of the diagram, there is an unstable stationary point~\cite{Pilatowsky2019}, which produces a one-point scar in addition to the scar associated to the blue orbit. The localization measure is larger in Fig.~\ref{fig01}~\panelref{b}{1} and \panelref{b}{2} simply because in these cases the phase space is very small. It is also larger for the states in the high energy region in Fig.~\ref{fig01}~\panelref{a}{5}, \panelref{a}{6}, \panelref{b}{4}, \panelref{b}{5}, \panelref{b}{6}, because they spread beyond the marked periodic orbits. As the energy increases and one approaches the chaotic region, more unstable periodic orbits emerge in the classical limit, enhancing the likelihood that a single quantum eigenstate gets scarred by different periodic orbits. We stress, however, that even for those high-energy states with larger values of $\mathfrak{L}_{k}$, the drawn periodic orbits cast a bright green shadow that is clearly visible in the Husimi projections.

It is important to make it clear that scarring and localization, despite related, are not synonyms. Of course, there is no eigenstate with a large value of $\mathfrak{L}_{k}$ that would at the same time have a high value of our scarring measure, but there is more to the relationship between these two concepts. A highly localized eigenstate is scarred by few periodic orbits of a particular family and therefore has a high value of the scarring measure for that family, although it has low values for the other families of periodic orbits~\cite{Pilatowsky2021}. Furthermore, even the most delocalized eigenstates are still scarred, but now by periodic orbits from different families~\cite{Revuelta2020}.

In Fig.~\ref{fig02}, we take a step further in the analysis of localization and scarring. In the large panel in Fig.~\ref{fig02}~(a), we show $\mathfrak{L}_{k}$ against energy for all eigenstates between $\epsilon_\text{GS}$ and $\epsilon=0.06$. This plot is equivalent to a Peres lattice~{\cite{Peres1984PRL} for expectation values of observables, as used in studies of chaos and thermalization. In the low-energy regular regime, $\mathfrak{L}_{k}$ is organized along lines that can be classified with quasi-integrals of motion linked with classical periodic orbits~\cite{Bastarrachea2017a}.  Conversely, as the system enters the chaotic region at higher energies, the distribution of $\mathfrak{L}_{k}$ becomes dense and looses any order. Notice, however, that all eigenstates in the chaotic region have values of $\mathfrak{L}_{k}$ much lower than  $1$, mostly clustering below $1/2$. 

The value $\mathfrak{L}\sim 1/2$ marks a limit on the spreading of any pure state in the high energy shells of the phase space. To show this, we build  random states $\ket*{\cal{R}_\epsilon}=\sum_k c_k \ket{E_k}$, where $c_k$ are complex random numbers from a Gaussian energy profile centered at energy $\epsilon$ (see Methods). The values of  the localization measure $\mathfrak{L}(\epsilon,\hat{\rho}_\epsilon)$ for four different random states $\hat{\rho}_\epsilon=\dyad*{\cal{R}_\epsilon}$ centered at increasing energies $\epsilon$ between $-0.6$ and $-0.1$ are given in Figs.~\ref{fig02}~\panelref{r}{1}-\panelref{r}{4}, and indeed $\mathfrak{L}\sim 1/2$. This upper bound on the phase-space delocalization is not due to quantum scarring, but to quantum interference effects. 

The panels in Figs.~\ref{fig02}~\panelref{r}{1}-\panelref{r}{4} display the projected Husimi distributions $\widetilde{\mathcal{Q}}_{\epsilon,\hat{\rho}_\epsilon}(Q,P)$ of the four random states and those in Fig.~\ref{fig02}~\panelref{s}{1}-\panelref{s}{22} show the distributions for 22 eigenstates taken at fixed steps of $k$ with $\epsilon_k \in [-0.62,0.06]$ (various other examples are provided in the Supplementary Note 2). The difference between random states and eigenstates is clear. The Husimi projections of the random states do not show structures that resemble closed periodic orbits, so they are not scarred, while the Husimi projections of all eigenstates in the chaotic region do show those structures. In contrast to Fig.~\ref{fig01} (see also~\cite{Pilatowsky2021}), we have not identified the periodic orbits associated with Fig.~\ref{fig02}~\panelref{s}{1}-\panelref{s}{22}, but the visible circular patterns are clear evidence of periodic orbits. Their existence is supported by the shape of the Husimi projections and by knowing the generic direction of the classical Hamiltonian flow. The patterns display all the features of periodic orbits: they always cross the line $P = 0$ perpendicularly, they are symmetric  along both the horizontal $Q$ and $P$ axes, and they visibly form closed loops. There is no quantum effect other than scarring that would produce such patterns. One therefore deduces that deep in the chaotic region of the Dicke model, all quantum eigenstates are scarred, although they have different degrees of scarring, as seen by the patterns, and they have different degrees of localization, ranging  from strong localization ($\mathfrak{L}\sim 0.1$) to states that are nearly as much delocalized as random states.

%%%%%%%%%%%%%%%%%%%%%%%%%%%%%%%%%%%%%%%%%%%%%%%%%%%%%
%%%%%%%%%%%%%%%%%%%% FIGURE 3 %%%%%%%%%%%%%%%%%%%%%%%%%%%
\begin{figure*}[ht]
\includegraphics[width=1\textwidth]{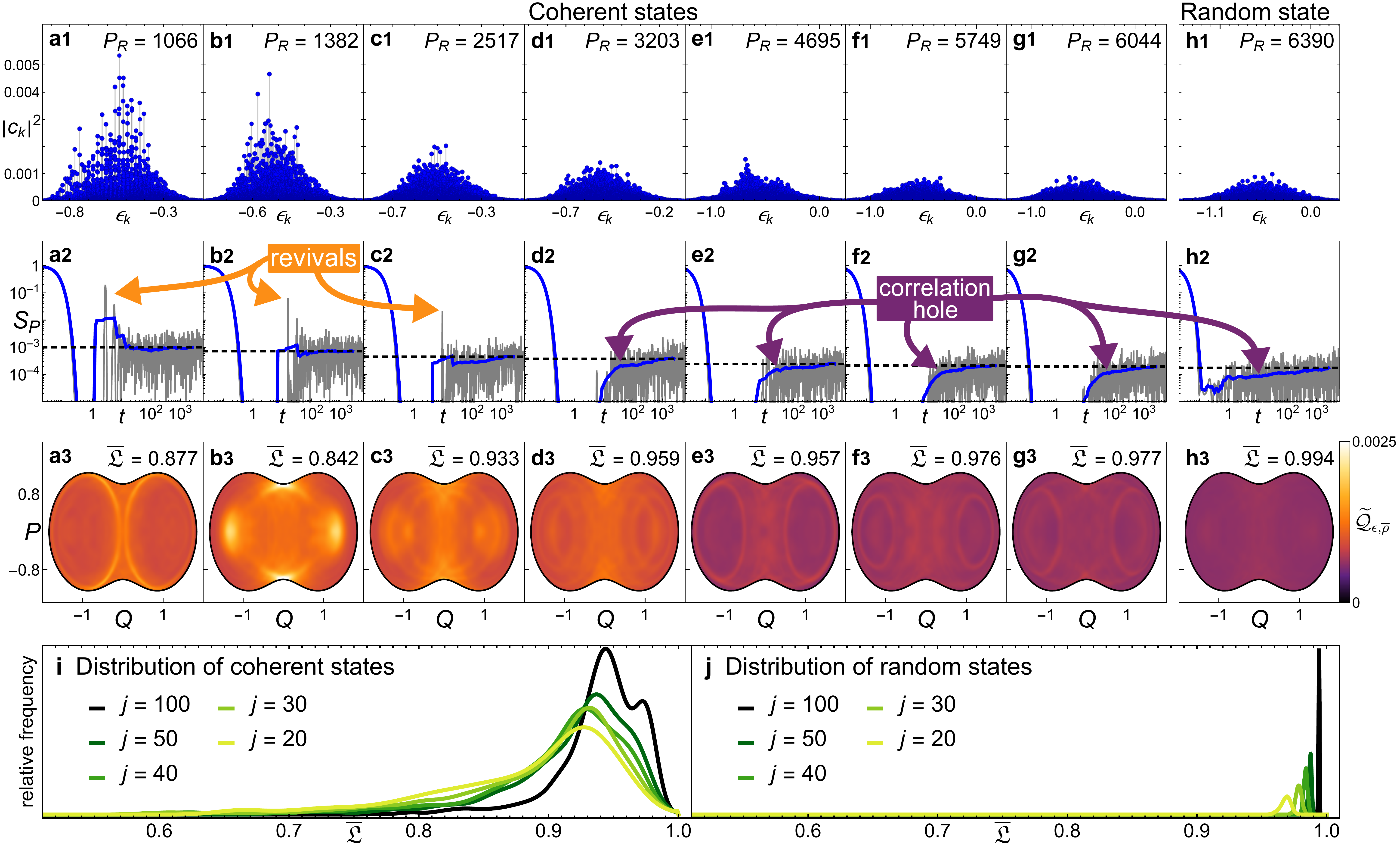} 
\caption{\textbf{Dynamical behavior of coherent and random states.}\textbf{ \panelref{a}{1}-\panelref{g}{1}} Energy distribution for coherent states with different values of $P_R$ and centered at $\epsilon=-0.5$, $j=100$. \textbf{\panelref{h}{1}} Energy distribution for a random state $\ket{{\cal R}_\epsilon}$ with energy width $\sigma=0.3$ centered at $\epsilon=-0.5$, $j=100$. \textbf{\panelref{a}{2}-\panelref{h}{2}} Quantum survival probability (gray solid line), its running average (blue solid line) and its asymptotic value (black dashed line) for the corresponding states of panels \panelref{a}{1}-\panelref{h}{1}. \textbf{\panelref{a}{3}-\panelref{h}{3}} Projected Husimi distributions of the time-averaged ensemble, $\widetilde{\mathcal{Q}}_{\epsilon=-0.5,\overline{\rho}}$ for the corresponding states of panels \panelref{a}{1}-\panelref{h}{1} with the values of $\overline{\mathfrak{L}}(\epsilon=-0.5,\hat{\rho})$ indicated. \textbf{(i)} Distribution of $\overline{\mathfrak{L}}(\epsilon,\hat{\rho})$ for a set of $1551$ coherent states that are evenly distributed along the atomic variables $(Q,P)$ at $\epsilon=-0.5$ and $p=0$. This is done for system sizes $j=20, 30, 40, 50, 100$, as indicated. \textbf{(j)} Distribution of $\overline{\mathfrak{L}}(\epsilon,\hat{\rho})$ for 500 random states with $\epsilon=-0.5$ and $\sigma=0.3$ for $j=20, 30, 40, 50, 100$.}
\label{fig03}
\end{figure*}
%%%%%%%%%%%%%%%%%%%%%%%%%%%%%%%%%%%%%%%%%%%%%%%%%%%%%
%%%%%%%%%%%%%%%%%%%%%%%%%%%%%%%%%%%%%%%%%%%%%%%%%%%%%

Our results are sharpened as one approaches the semiclassical limit. The patterns indicating scarring do not fade away as the system size increases. Quite the opposite, as $j=1/\hbar_\text{eff}$ increases, the periodic orbits get better defined in the Husimi projections (cf. the figures for $j=30$ and $j=100$ in the Supplementary Note 3). To study the dependence of the level of phase-space localization on system size, we show the distributions  of $\mathfrak{L}$ for random states  [Fig. \ref{fig02} (b)] and for eigenstates in the chaotic region [Fig. \ref{fig02} (c)]  for different values of $j$. For the random states, $\mathfrak{L}$ is concentrated around 1/2 and the width of the distribution decreases as $j$ increases, corroborating that $\mathfrak{L}=1/2$ is indeed the delocalization upper bound.  In contrast, for the eigenstates in the chaotic regime, the distributions are skewed and broader. The tail at small values of $\mathfrak{L}_k$ does not change as $j$ increases, showing that the highly localized states persist, while the portion of the states with large $\mathfrak{L}_k$ decreases, suggesting that for large system sizes, none of the eigenstates reach values of $\mathfrak{L}_k>1/2$. 

The ubiquitous scarring revealed by our studies motivates the question of whether scarring in other quantum models is also the rule.  We have found hints in the literature suggesting that our findings may actually be quite general. For example, in Ref.~\cite{Revuelta2020}, the authors reconstruct significant portions of the spectrum of a quantum chaotic system using only periodic orbits. This means that all of these eigenstates are described by those periodic orbits and are therefore scarred. In Ref.~\cite{Muller1994}, the authors claim that the great majority of the eigenstates of the hydrogen atom in a magnetic field may be related to periodic orbits, indicating that scars must be the rule. But to provide a definite answer, a phase-space analysis similar to the one presented here is needed.

\vskip 0.2 cm 
{\bf Quantum ergodicity.} We have so far discussed two concepts -- quantum scarring and phase-space localization -- that are related, but are not equal. How about their relationship with quantum ergodicity? In the classical limit, a system is ergodic if the trajectories cover the energy shell homogeneously. We then adopt the same definition for quantum ergodicity. To quantify how much of the energy shell is visited on average by the evolved state $\hat{\rho}(t)=e^{-i\hat{H}_{D}t}\hat{\rho}\, e^{i\hat{H}_{D}t}$, we consider the infinite-time average \cite{Sunada1997,Schnack1996},
\begin{equation}
\overline{\rho}=\lim_{T\to \infty} \frac{1}{T} \int_0^T \dif t \, \hat{\rho}(t),
\end{equation}
and compute $\overline{\mathfrak{L}}(\epsilon,\hat{\rho})\equiv\mathfrak{L}(\epsilon,\overline{\rho})$ with equation~\eqref{eqn:LocMeasDef}. If the whole energy shell at $\epsilon$ is homogeneously visited by $\hat{\rho}$, then $\overline{\mathfrak{L}}(\epsilon,\hat{\rho})=1$.  We thus say that a quantum state $\hat{\rho}$ is ergodic over the energy shell $\epsilon$ if  $\overline{\mathfrak{L}}(\epsilon,\hat{\rho})=1$. According to this definition, all stationary states in the chaotic region of the Dicke model are non-ergodic, since  $\overline{\mathfrak{L}}(\epsilon_k,\hat{\rho}_k)=\mathfrak{L}(\epsilon_k,\hat{\rho}_k)\lesssim 1/2$, as shown above.

How about non-stationary states, such as coherent states or random states, are they ergodic? We study the evolution of initial coherent states $|\Psi(0) \rangle = \ket{\bold{x}_0} = \sum_k c_k |E_k\rangle$ with mean energies $\epsilon=-0.5$, that are in the chaotic region (see Methods). We select both coherent states that are highly localized and  delocalized in the energy eigenbasis, with the degree of delocalization measured by the participation ratio $P_R= (\sum_k |c_k|^4)^{-1}$.  Their energy distributions are shown in Figs.~\ref{fig03}~\panelref{a}{1}-\panelref{g}{1}. The components of the states with low $P_R$ are bunched around specific energy levels [Fig.~\ref{fig03}~\panelref{a}{1} and \panelref{b}{1}], exhibiting the comb-like structure typical of scarred states \cite{Heller1984}. As $P_R$ increases, the coherent states become more spread in the energy eigenbasis, looking more similar to the random state $\ket{\Psi(0)}=\ket{{\cal R}_\epsilon}$ shown in Fig.~\ref{fig03}~\panelref{h}{1}, whose mean energy is also $\epsilon=-0.5$. 

The evolution of the survival probability, $S_P(t) = |\langle \Psi(0) |e^{-i\hat{H}_{D}t}|\Psi(0) \rangle|^2$, for the coherent states with low $P_R$ leads to large revivals before the saturation of the dynamics \cite{Alhambra2020}, as seen in Figs.~\ref{fig03}~ \panelref{a}{2}, \panelref{b}{2}, and \panelref{c}{2}. This contrasts with the evolution of the coherent states with large $P_R$, such as those in Figs.~\ref{fig03}~\panelref{d}{2}-\panelref{g}{2}, and the evolution of the random state in Fig.~\ref{fig03}~\panelref{h}{2}. In these cases,  the approach to the asymptotic value of $S_P(t)$ is much smoother and exhibits the so-called correlation hole, which corresponds to the ramp towards saturation. The correlation hole reflects the presence of correlated eigenvalues and is a quantum signature of classical chaos~\cite{Leviandier1986,Torres2017PTR,Villasenor2020}.

The values of $\overline{\mathfrak{L}}(\epsilon, \hat{\rho})$ for the states in Figs.~\ref{fig03}~\panelref{a}{1}-\panelref{h}{1} are indicated in Figs.~\ref{fig03}~\panelref{a}{3}-\panelref{h}{3}. The random state is indeed ergodic, reaching $\overline{\mathfrak{L}} \sim 1$. For the coherent states, $\overline{\mathfrak{L}}$ increases as $P_R$ does, but even for the states with the largest values of the participation ratio, $\overline{\mathfrak{L}}$ is still slightly under $1$. The analysis of the dependence of $\overline{\mathfrak{L}}$ on system size is done in Figs. \ref{fig03} (i)-(j). The distribution of $\overline{\mathfrak{L}}$ for random states [Fig. \ref{fig03}~(j)] gets narrower and better centered at $\overline{\mathfrak{L}}=1$ as $j$ increases, confirming that these states indeed behave ergodically. The distribution for the coherent states [Fig. \ref{fig03}~(h)] is much broader. The center moves towards larger values as $j$ increases, but it is not clear whether there will ever be a significant portion of the initial coherent states with $\overline{\mathfrak{L}} \sim 1$.

In Figs.~\ref{fig03}~\panelref{a}{3}-\panelref{h}{3}, we plot the projected time-averaged Husimi distributions $\widetilde{\mathcal{Q}}_{\epsilon,\overline{\rho}}(Q,P)$ for the states in Figs.~\ref{fig03}~\panelref{a}{1}-\panelref{h}{1}. Remarkably, even for the coherent states with high $P_R$, which do not exhibit any comb-like structure in their energy distributions and do not show revivals in the evolution of their survival probability, we still see an enhancement around unstable periodic orbits, as revealed by a careful inspection of Figs.~\ref{fig03}~\panelref{d}{3}-\panelref{g}{3} and in contrast with the absence of any pattern for the random state in Fig.~\ref{fig03}~\panelref{h}{3}. This unexpected manifestation of dynamical scarring can only be observed in phase space, having no identifiable signature in the energy distribution of the initial states or in the evolution of the survival probability. Thus, revivals in the long-time quantum dynamics are signs of a scarred initial state, but lack of revivals do not exclude the presence of scarring, it just indicates that if it exists it is at a low degree.

%%%%%%%%%%%%%% CONCLUSIONS %%%%%%%%%%%%%%%%
\vskip 0.2 cm
{\large \bf Discussion} 

The three main concepts investigated and compared in this work were quantum scarring, phase-space localization, and quantum ergodicity. We showed that for the Dicke model, all eigenstates in the chaotic region are scarred, although with different degrees of scarring and different levels of phase-space localization. Evidently, an eigenstate that is strongly scarred is also highly localized in phase space, but a single eigenstate may be scarred by different periodic orbits and reach levels of delocalization almost as high as a random state. We also showed that any pure state – even without any trace of scarring – is localized in phase space, and that ergodicity is an ensemble property, achievable only through temporal averages. Thus, scarring, localization and lack of ergodicity are not synonyms, although connections exist.

The ubiquitous scarring of the eigenstates does not immediately translate into the breaking of quantum ergodicity. All eigenstates are certainly non-ergodic, since they never reach complete delocalization in phase space, but if a non-stationary state visits on average the available phase-space homogeneously, then it is ergodic. Random states, for example, are ergodic. The analysis of initial coherent states showed that some are heavily scarred, resulting in the strong breaking of ergodicity that translates into the usual revivals of the survival probability. More interesting is the subtle behavior of the majority of the initial coherent states, which do not display revivals in the quantum dynamics or the comb-like structure in their energy distributions, but yet show some degree of scarring. 

Analyses that focus on the Hilbert space, such as the energy distribution of  the initial states or the fluctuations of eigenstate expectation values in Peres lattices and comparisons with thermodynamic averages, that are often done in studies of the eigenstate thermalization hypothesis (ETH), may miss the ubiquitous scarring observed in this work. For this feature to be revealed, one needs to look at the structures of the states in phase space.

Our results for the Dicke model are, of course, important, due to the widespread theoretical interest in this model and the fact that it is employed to describe experiments with trapped ions and superconducting circuits. But the repercussion of the findings discussed here goes beyond the limit of spin-boson systems by raising the question of whether scarring in other quantum systems is also the rule and not the exception. Our work provides the appropriate tools to address this question. The phase-space method that we developed is applicable to any quantum system that has a tractable phase-space. Whether these studies could eventually be extended also to interacting many-body quantum systems, such as interacting spin-1/2 models, will depend on the viability of their semiclassical analysis, and some recent works give reasons for optimism~\cite{Akila2017,Rammensee2018,SerbynARXIV}.

\newpage
%%%%%%%%%%%%%% METHODS %%%%%%%%%%%%%%%%%%%%%%%%%%%%%%
{\bf\large Methods}

{\bf Classical Hamiltonian.} To construct the classical Hamiltonian in a four-dimensional phase space $\mathcal{M}$ with coordinates $\bold{x}=(q,p;Q,P)$, we use the Glauber-Bloch coherent states $\ket{\bold{x}}=\ket{q,p}\otimes\ket{Q,P}$ \cite{Deaguiar1991,Deaguiar1992,Bastarrachea2014a,Bastarrachea2014b,Bastarrachea2015,Chavez2016}. They are tensor products of the bosonic Glauber coherent states 
$|q,p\rangle=e^{-(j/4)\left(q^{2}+p^{2}\right)}e^{\left[\sqrt{j/2}\left(q+ip\right)\right]\hat{a}^{\dagger}}|0\rangle$
and the pseudo-spin Bloch coherent states  $|Q,P\rangle=\left(1-\frac{Z^{2}}{4}\right)^{j}e^{\left[\left(Q+iP\right)/\sqrt{4-Z^{2}}\right]\hat{J}_{+}}|j,-j\rangle$, where $Z^{2}=Q^{2}+P^{2}$, $|0\rangle$ denotes the photon vacuum,   $|j,-j\rangle$ designates the state with all atoms in their ground state, and $\hat{J}_{+}$ is the raising atomic operator. The classical Hamiltonian is given by 
\begin{align}
\label{eqn:cla_hamiltonian_QP}
h_\text{cl}(\bold{x}) &=\frac{\omega}{2}(q^{2}+p^{2})
 +\frac{\omega_{0}}{2}Z^{2} +2\gamma Q q\sqrt{1-\frac{Z^{2}}{4}} -\omega_{0}.
\end{align}

{\bf Efficient basis and system sizes.}
The efficient basis is  the Dicke Hamiltonian \eqref{eqn:qua_hamiltonian} eigenbasis in the limit $\omega_{0}\rightarrow0$, which can be analytically  obtained   by a displacement of the bosonic operator $\hat{A}=\hat{a}+(2\gamma/(\omega\sqrt{\mathcal{N}}))\hat{J}_{x}$ and a rotation of $-\pi/2$ around the $y$ axis of the collective pseudo-spin operators $\left(\hat{J}_{x},\hat{J}_{y},\hat{J}_{z}\right)\rightarrow\left(\hat{J}'_{z},\hat{J}'_{y},-\hat{J}'_{x}\right)$, 
\begin{equation}
|N;j,m'\rangle=\frac{(\hat{A}^{\dagger})^{N}}{\sqrt{N!}}|N=0;j,m'\rangle,
\end{equation}
where $N$ is the eigenvalue of the modified bosonic number operator $\hat{A}^{\dagger}\hat{A}$ and $m'=m_{x}$ is the eigenvalue of the original collective pseudo-spin operator $\hat{J}_{x}$. The modified bosonic vacuum states in $|N=0;j,m'\rangle=|N=0\rangle\otimes |j,m'\rangle$ are  Glauber coherent states  $|N=0\rangle=|-2\gamma m'/(\omega\sqrt{\mathcal{N}})\rangle$. The Hilbert space dimension of this basis is given by $D=(2j+1)(N_{\text{max}}+1)$, where $N_{\text{max}}$ designates a cutoff of the modified bosonic subspace.

This basis allows to work with larger values of $j$ by reducing the value of $N_\text{max}$ required for convergence of the high-energy eigenstates. With $j=100$ and $N_{\text{max}}=300$ ($D=60501$), we are able to get  30825 converged eigenstates covering the whole energy spectrum up to $\epsilon=0.853$. Having converged eigenstates in such a high-energy regime would be infeasible with the usual Fock basis for $j=100$. \cite{Bastarrachea2011, Bastarrachea2014PSa, Bastarrachea2014PSb, Bastarrachea2016PRE}

{\bf Husimi projection and localization measure.} To compute the Husimi projection given in equation (\ref{eq:projQdef}) and the  localization measure given by equation (\ref{eqn:LocMeasDef}), one has to compute integrals of the form 
\begin{equation}
\widetilde{f}(Q,P)=\iint \dif q \dif p \, \delta(\epsilon - h_\text{cl}(\bold{x})) f(\bold{x}),
\end{equation}
where $\bold{x}=(q,p;Q,P)$ and $f(\bold{x})$ is a non-negative function in the phase space. For the localization measure, note that $\mathfrak{L}(\epsilon,\hat{\rho})^{-1}=\frac{1}{N}\iint \dif Q \dif P \widetilde{f_2}(Q,P)$ with $f_2=\mathcal{Q}^2_{\hat{\rho}}$, $N=(\iint \dif Q \dif P \widetilde{f_1}(Q,P))^2 / \mathcal{V}(\epsilon)$ with $f_1=\mathcal{Q}_{\hat{\rho}}$, and $\mathcal{V}(\epsilon)=\iint \dif Q \dif P \widetilde{f_0}(Q,P)$ with $f_0(\bold{x})=1$.

By using properties of the $\delta$ function, those integrals are reduced to
\begin{equation}
\label{eqn:NumHu}
\widetilde{f}(Q,P) = \int_{p_{-}}^{p_{+}}  \dif p \frac{\sum_{q_{\pm}}f(q_{\pm},p;Q,P)}{\sqrt{\Delta(\epsilon,p,Q,P)}} ,
\end{equation}
where $q_{\pm}$ are the two solutions in $q$ of the second-degree equation $h_\text{cl}(q,p;Q,P)=\epsilon$,
\begin{align}
\Delta(\epsilon,p,Q,P) & = \abs{\frac{\partial h_\text{cl}}{\partial q} (q_\pm,p;Q,P)}^{2} \nonumber \\
& = 2 \omega \omega_{0} \left(\frac{\epsilon}{\omega_{0}} + 1 - \frac{Q^{2} + P^{2}}{2}\right) + \nonumber \\ 
& 4 \gamma^{2} Q^{2} \left(1 - \frac{Q^{2} + P^{2}}{4}\right) - \omega^{2} p^{2},
\end{align}
and $p_{\pm}$ are the two solutions in $p$ of the second-degree equation $\Delta(\epsilon,p,Q,P)=0$. Because of the form of the weight $1/\sqrt{\Delta}$, the integral given by equation \eqref{eqn:NumHu} may be computed efficiently using a Chebyshev-Gauss quadrature method. 

 It is worth noting that a quantum state with a Wigner distribution  that is constant in an energy shell will lead to a Husimi function that  needs not to be constant within the same energy shell. This is because the Husimi distribution is the convolution of the Wigner distribution with the Gaussian Wigner distributions of the coherent states, which have different energy widths due to the geometry of the energy shells in the phase space. This rather marginal effect may be seen in Fig. \ref{fig03} \panelref{h}{3}, where there is a barely visible weak concentration towards the center of the plot causing $\overline{\mathfrak{L}}$ to be slightly under $1$. We stress that this effect is not related to quantum scarring. It is just a manifestation of the phase-space geometry in the Husimi distributions. 
 
{\bf Coherent states.} The coherent states $\ket{\bold{x}}$ are described by four coordinates $\bold{x} = (q,p;Q,P)$. To select coherent states at a given energy $\epsilon$, we solve the second-degree equation 
$h_\text{cl}(q,p;Q,P)=\epsilon$ for $q$, which yields two solutions $q_+(\epsilon,p,Q,P)\geq q_-(\epsilon,p,Q,P)$ \cite{Villasenor2020}. All of the initial coherent states shown in Fig. \ref{fig03} have bosonic variables given by  $q=q_+(\epsilon=-0.5,p,Q,P)$ and $p=0$. The coherent states shown in  Fig. \ref{fig03} \panelref{a}{1}-\panelref{g}{1} have atomic coordinates given by $(Q,P)= (1.75,0)$~\panelref{a}{1}, $(0.5,0.75)$~\panelref{b}{1}, $(0.75,0.5)$~\panelref{c}{1}, $(1.25,0.25)$~\panelref{d}{1}, $(-1.25,1)$~\panelref{e}{1}, $(-1.25,0.75)$~\panelref{f}{1}, and  $(-0.75,0.5)$~\panelref{g}{1}. The atomic coordinates of the $1551$ coherent states whose distributions of $\overline{\mathfrak{L}}$ are shown in Fig~\ref{fig03}~(i) were selected by constructing a rectangular grid with step $\Delta Q=\Delta P =0.05$ from $(Q_i,P_i)= (-2,-2)$ to $(Q_f,P_f)=(2,0)$. Of the $3321$ points inside of this grid, $1551$ have allowed values of $Q,P$ (i.e. they satisfy $Q^2 + P^2 \leq 4$) and fall inside of the energy shell at $\epsilon=-0.5$ (i.e. there exists $q_+(\epsilon=-0.5,p=0,Q,P)$). We use these $1551$ coherent states to compute the distributions in Fig~\ref{fig03}~(i).

{\bf Random states.} The state $\ket*{\cal{R}_\epsilon}=\sum_k c_k \ket{E_k}$ is built by sampling random numbers $r_k>0$ from an exponential distribution $\lambda e^{-\lambda x}$ and random phases $\theta_k$ from a uniform distribution in $[0,2\pi)$. We use
\begin{equation}
c_k=\sqrt{\frac{r_k \rho(E_k)}{\nu(E_k)M}}\, e^{i \theta_k}
\end{equation}
where $\nu(E)$ is the density of states, $\rho(E)$ is a Gaussian profile of width $j \sigma$ centered at energy $j \epsilon$, and $M$ ensures normalization. This way, $\ket*{\cal{R}_\epsilon}$ has a defined energy center $\epsilon$, where the Husimi projection and the localization measure are calculated. \cite{Lerma2019, Villasenor2020}
%%%%%%%%%%%%%% ACKNOWLEDGMENTS and MORE %%%%%%%%%%%%%%%%
\vskip 0.3 cm 
{\bf\large Data availability}\\
All the data that support the plots within this paper and other findings of this study are available from the corresponding authors upon request.

\vskip 0.3 cm 
{\bf\large Code availability}\\
All the computational codes that were used to generate the data presented in this paper are available from the corresponding authors upon request.

\vskip 0.3 cm 
{\bf\large Acknowledgments} \\
We thank D. Wisniacki for his valuable comments and acknowledge the support of the Computation Center - ICN, in particular of Enrique Palacios, Luciano D\'iaz, and Eduardo Murrieta.  SP-C, DV and JGH acknowledge financial support from the DGAPA- UNAM project  IN104020, and SL-H from the Mexican CONACyT project CB2015-01/255702. LFS was supported by the NSF grant No. DMR-1936006.

\vskip 0.3 cm 
{\bf\large Author contributions}\\
SP-C and DV were responsible for most of the calculations and the development of the work. MAB-M, SL-H, LFS and JGH provided the original ideas and shaped the manuscript.

\vskip 0.3 cm 
{\bf\large Competing interests}\\
The authors declare no competing interests.

\vskip 0.3 cm 
{\bf\large Additional information}\\
{Supplementary information} is available for this paper.

%----------------------------------------------------------------------------------------
%	REFERENCE LIST
%----------------------------------------------------------------------------------------
%%%%%%%%%%%%%%%%%%%%%%%%%%%%%%%%%%%%%%%%%%%%%%%%%%
%%%%%% \bibliography{bibliography}

%%%%%%%%%%%%%%%%%%%%%%%%%%%%%%%%%%%%%%%%%%%%%%%%%%
\end{document}

\title{\Large Ubiquitous quantum scarring does not prevent ergodicity\\ Supplementary Information}

\author{Sa\'ul Pilatowsky-Cameo}
\author{David Villase\~nor}
\affiliation{Instituto de Ciencias Nucleares, Universidad Nacional Aut\'onoma de M\'exico, Apdo. Postal 70-543, C.P. 04510, CDMX, Mexico}
\author{Miguel~A.~Bastarrachea-Magnani}
\affiliation{Department of Physics and Astronomy, Aarhus University, Ny Munkegade, DK-8000 Aarhus C, Denmark}
\affiliation{Departamento de F\'isica, Universidad Aut\'onoma Metropolitana-Iztapalapa, San Rafael Atlixco 186, C.P. 09340, CDMX, Mexico}
\author{Sergio Lerma-Hern\'andez}
\affiliation{Facultad de F\'isica, Universidad Veracruzana, Circuito Aguirre Beltr\'an s/n, Xalapa, Veracruz 91000, Mexico}
\author{Lea~F.~Santos} 
\affiliation{Department of Physics, Yeshiva University, New York, New York 10016, USA}
\author{Jorge G. Hirsch}
\affiliation{Instituto de Ciencias Nucleares, Universidad Nacional Aut\'onoma de M\'exico, Apdo. Postal 70-543, C.P. 04510, CDMX, Mexico}

%%%%%%%%%%%%%%%%%%%%%%%%%%%%%%%%%%%%%%%%
\maketitle
%%%%%%%%%%%%%%%%%%%%%%%%%%%%%%%%%%%%%%%%

\subsection*{\large	Supplementary Note 1: \\Classical and quantum chaos in the Dicke model}

The Dicke model displays regular and chaotic behavior~\cite{Lewenkopf1991,Emary2003PRL,Emary2003,Bastarrachea2014b,Bastarrachea2015,Bastarrachea2016PRE,Chavez2016,Chavez2019}. For the parameters selected in the main text ($\omega=\omega_0=1, \gamma=2 \gamma_c$), the dynamics are regular up to $\epsilon\approx-1.6$ \cite{Chavez2016}, then there is a mixed region of regularity and chaos up to  $\epsilon\approx-0.8$, after which strong chaos sets in.
The onset of chaos is illustrated in Supplementary Fig.~\ref{fig01} for the classical limit (a)-(b) and for the quantum domain (c)-(d). 

Supplementary Fig.~\ref{fig01}~(a) shows the percentage of chaos defined as the ratio of
the number of chaotic initial conditions, determined by the Lyapunov exponent, over the total number of initial conditions for a very large sample. The percentage is presented as a function of the rescaled energy $\epsilon$ and the coupling strength $\gamma$. Following the vertical red dashed line marked at  $\gamma=2\gamma_c$, one sees that energies $\epsilon \sim -0.5$ are already deep in the chaotic region (light color). This is confirmed in Supplementary Fig.~\ref{fig01}~(b), where the Poincar\'e section for $\epsilon = -0.5$ exhibits hard chaos, that is, all chaotic trajectories cover the entire energy shell densely and have the same positive Lyapunov exponent.

Supplementary Fig.~\ref{fig01}~(c) displays the distribution $P(s)$ of the spacings $s$ between nearest-neighboring unfolded energy levels. The eigenvalues of quantum systems whose classical counterparts are chaotic are correlated and repel each other. In this case, $P(s)$ follows the Wigner surmise~\cite{Guhr1998}, as indeed seen in Supplementary Fig.~\ref{fig01}~(c). 

In Supplementary Fig.~\ref{fig01}~(d), we show the quantum survival probability, $S_P(t)=|\langle {\cal R}_\epsilon|e^{-i\hat{H}_{D}t}|{\cal R}_\epsilon\rangle |^2$  for a Gaussian ensemble of random initial states $|{\cal R}_\epsilon\rangle=\sum_k c_k \ket{E_k}$ whose components $|c_{k}|^{2}$ were generated through a random sampling (see Methods) and are centered at energy $\epsilon = -0.5$ in the chaotic region~\cite{Lerma2019, Villasenor2020}. The survival probability of individual random states are shown with gray solid lines, their ensemble average with an orange solid line, the running time average  with a blue solid line, which overlaps with a green line that represents an analytical curve derived from the Gaussian orthogonal ensemble of the random matrix theory~\cite{Lerma2019,Villasenor2020}. The asymptotic value of $S_P(t)$ is shown with a horizontal red dashed line. The green and blue curves exhibit a dip below the saturation value of the quantum survival probability known as correlation hole, which is a dynamical manifestation of spectral correlations. It contains more information than the level spacing distribution $P(s)$, since in addition to short-range correlations, it captures also long-range correlations~\cite{Leviandier1986,Wilkie1991,Alhassid1992,Torres2017PTR,Lerma2019}. We verified that most coherent states from the chaotic region develops the correlation hole. Exceptions to this pattern are the states very close to unstable periodic orbits of relatively short periods~\cite{Villasenor2020}.

The four panels of Supplementary Fig.~\ref{fig01} leave no doubt that the Dicke model reaches a limit of very strong chaos. This is the region for which our analysis of the quantum scars is developed.

\clearpage
\vspace*{2em}
%%%%%%%%%%%%%%%%%%%%%%%%%%%%%%%%%%%%%%%%%%%%%%%%%%%%%
%%%%%%%%%%%%%%%%%%%% FIGURE 1 %%%%%%%%%%%%%%%%%%%%%%%%%%%
\begin{center}
\includegraphics[width=0.6\columnwidth]{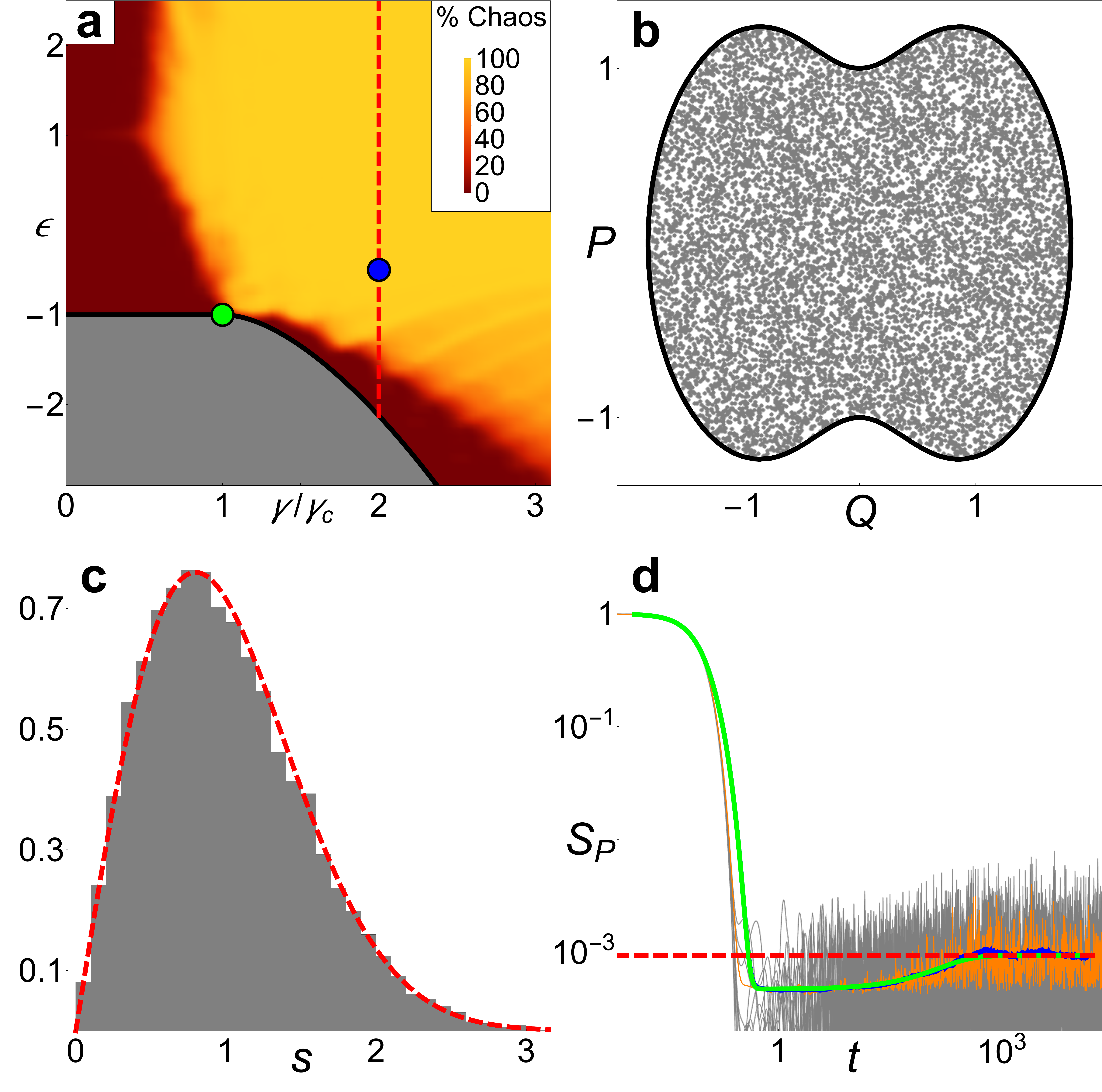}
\captionof{figure}{\textbf{Indicators of classical and quantum chaos.} \textbf{(a)} Percentage of chaos over classical energy shells. The black solid line follows the ground-state energy and the vertical red dashed line marks the coupling $\gamma=2\gamma_c$ chosen for our studies. The green dot marks the separation between the normal and the superradiant phase for the ground state. The blue dot represents the energy $\epsilon = -0.5$ used in the indicators (b) and (d). \textbf{(b)} Poincar\'e section $(p = 0)$ for the rescaled classical Hamiltonian $h_{\text{cl}}$ at energy $\epsilon = -0.5$. \textbf{(c)} Level spacing distribution of the unfolded spectrum (shaded area) for $22458$ levels in the energy region $\epsilon \in (-1,1.755)$ and Wigner surmise (red dashed line), $j=100$. \textbf{(d)} Survival probability for an ensemble of 500 random states (gray solid lines) centered at energy $\epsilon = -0.5$, ensemble average (orange solid line), running average (blue solid line), analytical curve from the random matrix theory (green solid line), and the asymptotic value (horizontal red dashed line)  (${j=100}$). 
}
\label{fig01}
\end{center}
%%%%%%%%%%%%%%%%%%%%%%%%%%%%%%%%%%%%%%%%%%%%%%%%%%%%%
%%%%%%%%%%%%%%%%%%%% FIGURE 1 %%%%%%%%%%%%%%%%%%%%%%%%%%%
\subsection*{\large Supplementary Note 2:\\ All eigenstates exhibit scars}

Supplementary Fig.~\ref{figSM2} shows the Husimi distributions of 160 eigenstates projected over the $(Q,P)$ plane for $j=100$. The eigenstates are selected from a list of 16,000 eigenstates with energies between $\epsilon_\text{GS}=-2.125$ and $\epsilon=0$, sampled in steps of 100 from $k=100$ to $k=16000$. The values of the localization measure $\mathfrak{L}_{k}$ are indicated in the panels. We select $\gamma=2\gamma_c$, so all these eigenstates are located in the red dashed line of Supplementary Fig.~\ref{fig01} (a). States with $k\leq 800$ ($\epsilon_k\leq -1.6$) are in the regular region, those with $800< k\leq 5600$ ($-1.6 < \epsilon_k\leq -0.82$) in the mixed region, and those with $k>5600$ ($ \epsilon_k> -0.82$) are in the region of strong chaos. In all projections, the Husimi distributions display ellipsoidal shapes that can be associated with periodic orbits in the classical limit once they are identified. 

\clearpage

%%%%%%%%%%%%%%%%%%%%%%%%%%%%%%%%%%%%%%%%%%%%%%%%%%%%%
%%%%%%%%%%%%%%%%%%%% FIGURE 2 %%%%%%%%%%%%%%%%%%%%%%%%%%%
\begin{center}
\centering
\includegraphics[width=1.0\textwidth]{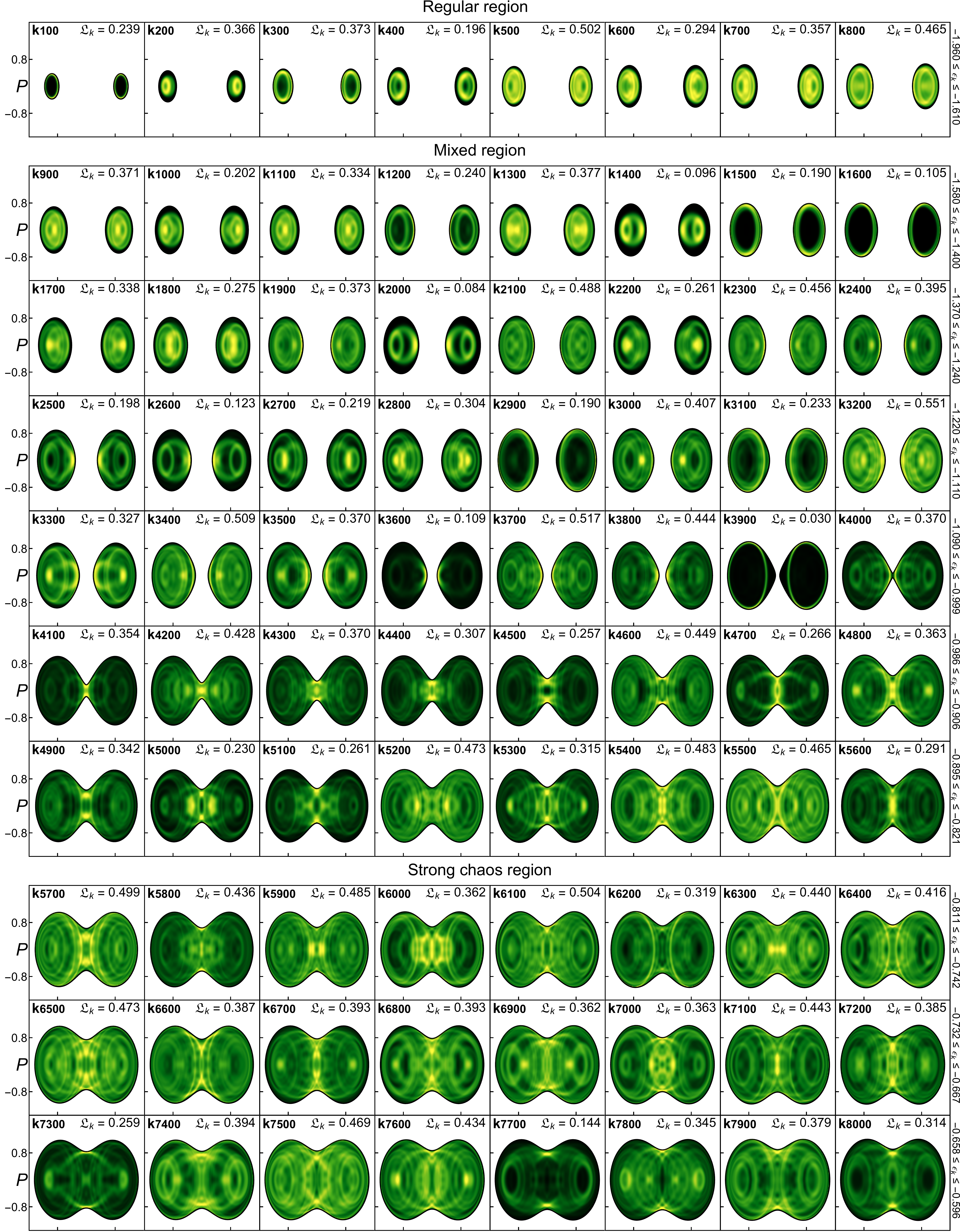}
{\footnotesize Figure continues on the next page}
\end{center}

\begin{center}
\includegraphics[width=1.0\textwidth]{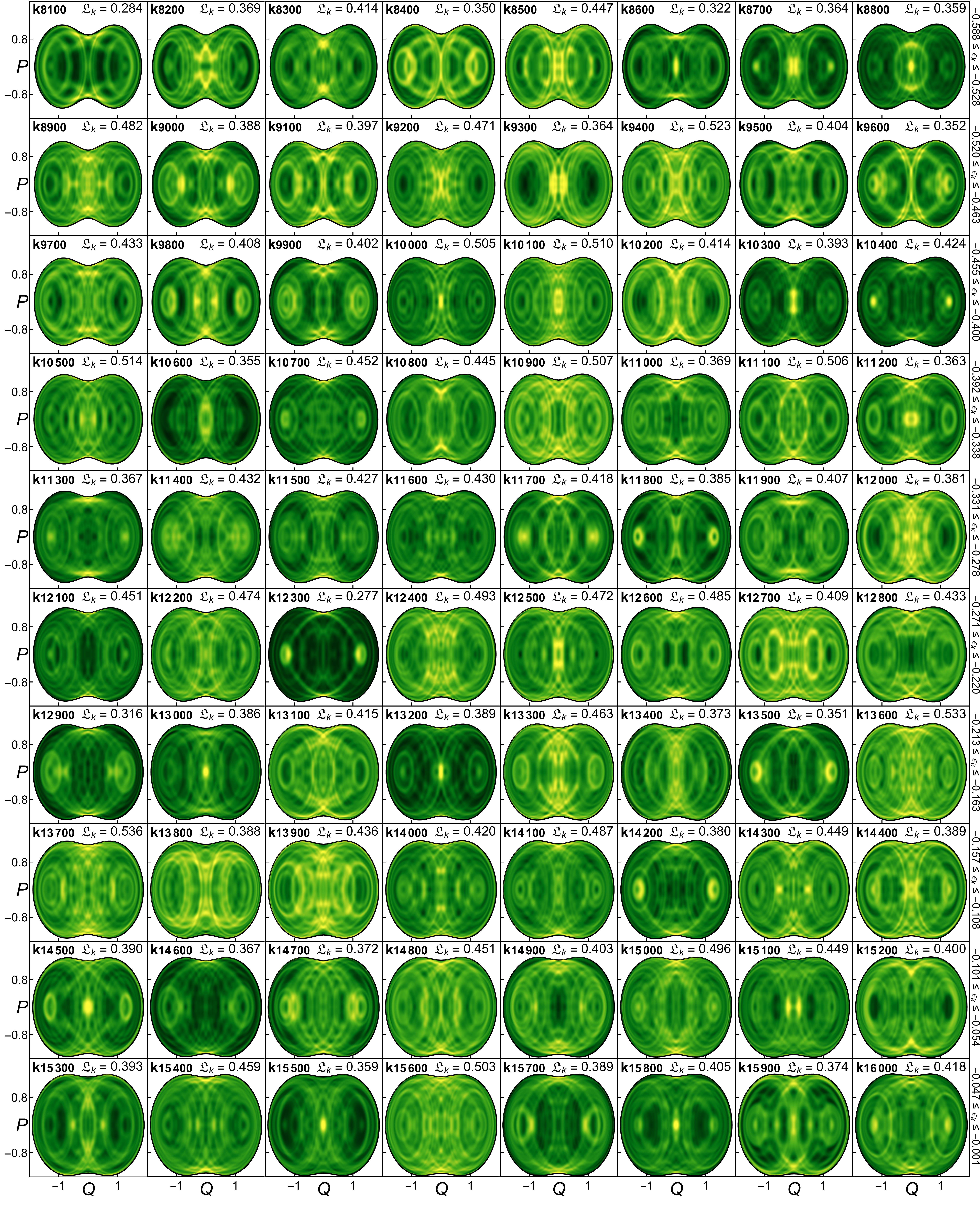}
\vspace{-2em}
\captionof{figure}{\textbf{Scars in all eigenstates.} Husimi projections $\widetilde{\cal{Q}}_k$ of 160 eigenstates  for  $j=100$. The values of $k$ are indicated in the top left of each panel, along with the value of $\mathfrak{L}_k$. The energy range is indicated on the right side of each row of panels. Lighter colors indicate higher concentrations, while black corresponds to zero.
}
\label{figSM2}
\end{center}
%%%%%%%%%%%%%%%%%%%%%%%%%%%%%%%%%%%%%%%%%%%%%%%%%%%%%
%%%%%%%%%%%%%%%%%%%% FIGURE 2 %%%%%%%%%%%%%%%%%%%%%%%%%%%

\subsection*{\large Supplementary Note 3:\\ Dependence on system size}
The Husimi distributions for some representative eigenstates are shown in Supplementary Fig.~\ref{figSM3} for $j=30$ and $j=100$. 
The patterns marking the periodic orbits are very similar. For each column, compare the top state ($j=30$) with the bottom state ($j=100$). They have very similar patterns, but the lines become better defined as $j$ increases. As the system size increases, more lines also appear, because more periodic orbits scar the states.
\\
\begin{center}
\includegraphics[width=1.0\textwidth]{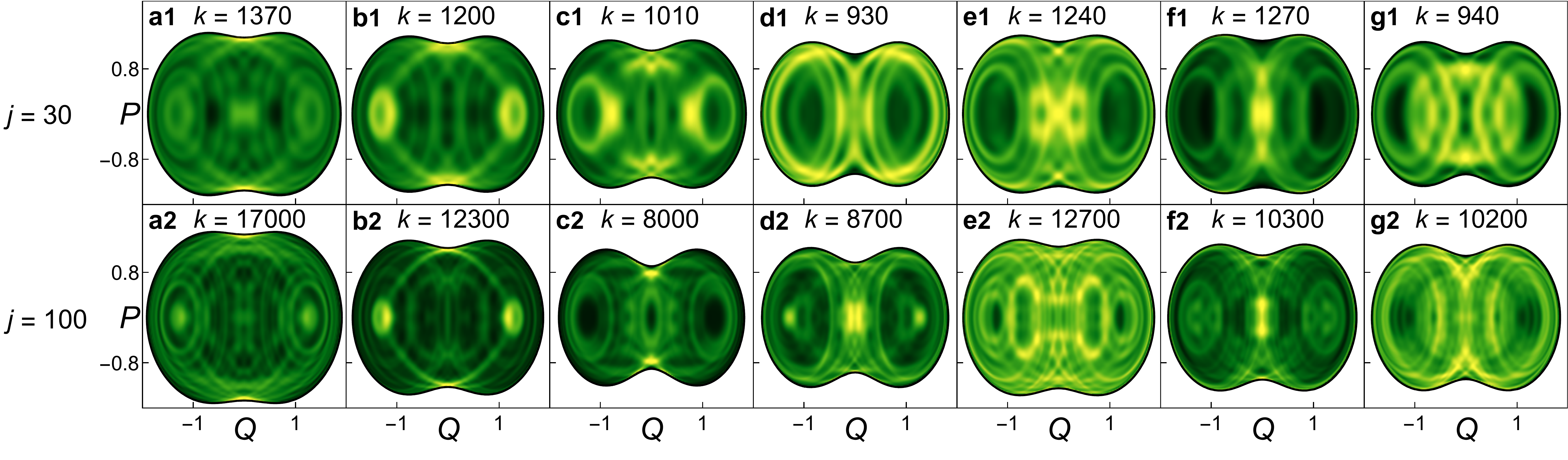}
\vspace*{-20px}
\captionof{figure}{\textbf{Husimi projections vs. system size.} Husimi projections $\widetilde{\cal{Q}}_k$ of 7 eigenstates  for  $j=30$ (first row) and $j=100$ (second row). Lighter colors indicate higher concentrations, while black corresponds to zero. The lines marking the periodic orbits become better defined as the system size increases.
}
\label{figSM3}
\end{center}
\vspace*{-15px}
\renewcommand{\baselinestretch}{1.08} 

%%%%%%%%%%%%%%%%%%%%%%%%%%%%%%%%%%%%%%%%%%%%%%%%%%
%\bibliography{../../bibliography}

%%%%%%%%%%%%%%%%%%%%%%%%%%%%%%%%%%%%%%%%%%%%%%%%%%